\documentclass[fleqn,usenatbib]{mnras}

\usepackage{newtxtext,newtxmath}
\usepackage[T1]{fontenc}

\DeclareRobustCommand{\VAN}[3]{#2}
\let\VANthebibliography\thebibliography
\def\thebibliography{\DeclareRobustCommand{\VAN}[3]{##3}\VANthebibliography}

\usepackage{graphicx}	% Including figure files
\usepackage{amsmath}	% Advanced maths commands
\usepackage{booktabs}
\usepackage{multirow}
\usepackage{subcaption}

\newcommand*    \msun{{\,\rm{M}_{\odot}}}

\newcommand*    \pc{{\,\mathrm{pc}}}

\newcommand*    \myr{{\,\rm Myr}}
\newcommand*    \gyr{{\,\rm Gyr}}
\newcommand*    \massb{M_{\rm b}}
\newcommand*    \massp{M_{\rm p}}
\newcommand*    \velb{V_{\rm b}}
\newcommand*    \enb{E_{\rm b}}
\newcommand*    \np{n_{\rm p}}

\title[Massive Perturbers BHB]{Perturber-Driven Dynamics of Supermassive Black Hole Binaries in Galaxy Mergers}

\author[Julian Chan et al.]{
Julian Chan$^{1}$\thanks{E-mail: julianjunyen.chan@surrey.ac.uk},
Alessia Gualandris$^{1}$,
Walter Dehnen$^{2}$ and
Justin I. Read$^{1}$
\\
$^{1}$School of Mathematics and Physics, Faculty of Engineering and Physical Sciences,
University of Surrey, Guildford GU2 7XH, UK\\
$^{2}$Astronomisches Rechen-Institut, Zentrum f{\"u}r Astronomie der Universit{\"a}t Heidelberg, M{\"o}nchhofstra\ss{}e 12-14, 69120, Heidelberg, Germany
}

\date{Accepted XXX. Received YYY; in original form ZZZ}

\pubyear{\the\year{}}

\begin{document}
\label{firstpage}
\pagerange{\pageref{firstpage}--\pageref{lastpage}}
\maketitle

\begin{abstract}
The orbital eccentricity of massive black hole binaries (MBHBs) at binary formation shapes the stochastic gravitational-wave background (GWB) detectable by pulsar timing arrays (PTAs). Previous $N$-body simulations show large run-to-run scatter in this quantity, dominated by Poisson noise, raising the question of whether physical substructure adds genuine astrophysical stochasticity. We test this with high-resolution re-simulations of a major merger from IllustrisTNG100-1, evolved with the \texttt{Griffin} $N$-body code. A no-perturber control is compared with two matched suites in which $f_{\rm target}=0.1$ of the primary bulge mass is redistributed into equal-mass perturbers of $10^7\msun$ ($\mu_{\rm p}\approx 3.2\times10^{-3}$) and $10^8\msun$ ($\mu_{\rm p}\approx 3.2\times10^{-2}$), with four realisations per scenario. The control gives $\sigma_e \approx 0.11$, consistent with the Poisson noise floor at this resolution. The $10^7\msun$ case gives $\sigma_e \approx 0.115$, indistinguishable from the control in its scatter, whereas the $10^8\msun$ case gives $\sigma_e \approx 0.26$, a factor of $2.4$ above the floor, though statistically marginal given only four realisations. This excess scatter coincides with larger event-aligned residuals in orbital energy and angular momentum and stronger torque spikes, consistent with near-impulsive perturber--MBHB encounters. In binary--single scattering theory the transition is set by the perturber--MBHB mass ratio $\mu_{\rm p}$: the $10^7\msun$ case remains diffusive while the $10^8\msun$ case approaches the near-impulsive regime. As the expected perturber population in massive ellipticals lies mostly below this regime, perturber-driven eccentricity randomisation is unlikely to matter for GWB-relevant MBHB mergers.
\end{abstract}


%%%%%%%%%%%%%%%%%%%%%%%%%%%%%%%%%%%%%%%%%%%%%%%%%%

%%%%%%%%%%%%%%%%% BODY OF PAPER %%%%%%%%%%%%%%%%%%

\section{Introduction}
Massive black hole binaries (MBHBs) formed in galaxy mergers are among the loudest sources of low-frequency (nHz) gravitational waves and are expected to dominate the stochastic gravitational-wave background (GWB) now detected by pulsar timing arrays (PTAs)~\citep{agazie_2023,antoniadis_2023,EPTA2024,reardon_2023,xu_2023}; see also~\citep{burke_2019}. The binaries that dominate the nHz background have total masses in the range $\sim 10^{8}$--$10^{10}\msun$~\citep{sesana_2013,agazie_2023}. Their imprint on the GWB depends sensitively on the orbital eccentricity at binary formation and during the subsequent hardening phase, as interactions with the surrounding galactic nucleus regulate both the rate at which MBHBs traverse the PTA band and how GW power is distributed over higher harmonics, which shape the strain spectrum of the GWB~\citep{sesana_2013,gualandris2022,fastidio2024}. High-resolution $N$-body simulations of galaxy mergers have shown that the eccentricity at binary formation is highly sensitive to small perturbations and exhibits substantial stochastic scatter among different realisations, particularly for nearly radial encounters~\citep{gualandris2022,rawlings2023,gualandris2026converging}. In previous resolution studies, this scatter decreases with increasing resolution approximately as $\sim 1/\sqrt{N}$, in line with Poisson-type sampling noise, and has since been demonstrated to be numerical in origin~\citep{gualandris2026converging}.

Yet these same simulations reveal that even small perturbations in the stellar background can drive substantial variations in eccentricity at binary formation, raising the question of whether analogous physical fluctuations, such as those sourced by massive perturbers (e.g.\ globular clusters, giant molecular clouds, or intermediate-mass black holes (IMBHs)), could induce genuine astrophysical stochasticity in the formation eccentricity.

\citet{rawlings2023} showed that, for nearly radial merger orbits, the binary eccentricity at the hard-binary separation is a non-linear function of the deflection angle during the final close SMBH encounter: small, parsec-scale perturbations to the plunging trajectory can shift the eccentricity across almost the full range $e \in [0,1]$. They argued that physical substructure in realistic galactic nuclei (nuclear gas, dense stellar clusters, and massive compact objects) should be sufficient to make the formation eccentricity effectively random, independently of resolution. Two physical pathways can produce such stochasticity: a single massive perturber delivering an impulsive kick large enough to appreciably alter the deflection angle of the final close encounter, or a population of lower-mass perturbers accumulating smaller deflections throughout the inspiral via a diffusive process. Whether either pathway produces a measurable effect depends on the perturber mass scale relative to the binary, motivating an $N$-body investigation of the transition between these regimes. Semi-analytical work has shown that massive perturbers can significantly shorten the relaxation time in galactic nuclei compared to two-body relaxation by stars alone \citep{Perets2007}, enhance loss-cone refilling and accelerate hardening near MBHs~\citep{PeretsAlexander2008,ArcaSedda2019}, but an $N$-body assessment of their impact on eccentricity at binary formation, and of the perturber mass scale required to produce a measurable physical effect, has not yet been carried out. In this paper, we present the first $N$-body investigation, using high-resolution merger simulations, testing two perturber mass regimes ($10^{7}$ and $10^{8}\msun$) that span the transition from the diffusive to the near-impulsive scattering regime.

This paper is organised as follows. Section~\ref{sec:theory} presents the theoretical framework for perturber-driven eccentricity diffusion and the diffusive--impulsive regime transition. Section~\ref{sec:methodology} describes the $N$-body models, perturber setup, and binary formation criterion. The results are presented in Section~\ref{sec:results} and discussed in Section~\ref{sec:discussion}. We summarise our conclusions in Section~\ref{sec:conclusions}.

\section{Theoretical framework}\label{sec:theory}
We estimate the expected effect of a population of massive perturbers with individual mass $\massp$ and number density $\np$ on a MBHB of mass $\massb$, moving with velocity $\velb \sim \sqrt{G\massb /a}$, where $a$ is its semi-major axis. This is the result of a sequence of interactions between
the MBHB and a passing massive perturber, which can be well described via binary–single scattering theory. Numerical scattering experiments show that, for hard binaries, the typical fractional change in binding energy per encounter is of order unity for comparable-mass intruders, while for unequal-mass encounters it scales with the ratio of the perturber mass to the binary mass \citep{Heggie1975, Hut1983, HeggieHut2003}. This can be expressed as:
\begin{equation}
	\frac{ \langle \Delta E \rangle}{\enb} \sim \xi  \frac{\massp}{\massb} ,
\end{equation}
where $\enb$ is the binary binding energy and $\xi \sim \mathcal{O}(0.1\text{--}1)$ is a dimensionless coefficient that depends weakly on the encounter velocity and geometry. This scaling reflects the fact that the energy available for exchange during an encounter is set by the kinetic energy of the incoming object, while the response of the binary is governed by its binding energy. In the limit of hard binaries, where the internal orbital velocity exceeds the velocity dispersion of the background, this leads to the well-known result that binaries tend to harden through repeated encounters \citep{Heggie1975}.
%For a three-body encounter with impact parameter $b$ and using the impulse approximation, the change in is given by
%\begin{equation}
%    \frac{\Delta E}{\enb} \sim \frac{\massp}{\massb} \left(\frac{\velb}{V}\right)^2 \left(\frac{a}{b}\right)^2
%\end{equation}

The angular momentum of a binary is generally more susceptible to perturbations than its binding energy, as even relatively distant encounters can induce significant torques. In the impulse approximation, a perturber of mass $\massp$ passing with velocity $V$ at impact parameter $b$ imparts a velocity kick $\Delta V\sim G \massp/(bV)$, leading to a fractional change in the binary angular momentum of order:
\begin{equation}\label{eq:angmom_kick}
	\frac{\Delta J}{J} \sim \frac{\massp}{\massb} \left(\frac{a}{b}\right) \left(\frac{\velb}{V}\right).
\end{equation}
Compared to energy changes, which scale as $(a/b)^2$, angular momentum perturbations are less strongly suppressed at large impact parameters, implying that a broader range of encounters contributes to its evolution. This behaviour underlies the known result that angular momentum diffuses more rapidly than energy in gravitational systems \citep{FrankRees1976, LightmanShapiro1977}.
As a consequence, stochastic fluctuations in the orbital angular momentum, and hence in the eccentricity, can arise even when the cumulative change in binding energy is small.

The cumulative effect of repeated perturbations can be described in terms of a diffusion process in eccentricity, following the orbit-averaged Fokker--Planck treatment of stellar encounters \citep{LightmanShapiro1977, CohnKulsrud1978, HeggieHut2003, BinneyTremaine2008, Merritt2013}. Using the relation between eccentricity and angular momentum, $e^2 = 1 - J^2/J_c^2$, where $J_c$ is the angular momentum of a circular orbit with the same energy, small perturbations give:
\begin{equation}
	\Delta e \sim \left[\frac{1-e^2}{e}\right] \frac{\Delta J}{J}.
\end{equation}

The eccentricity diffusion coefficient can then be written as:
\begin{equation}
	D_e = \frac{\langle (\Delta e)^2 \rangle}{ \Delta t} \sim \Gamma \langle (\Delta e)^2 \rangle,
\end{equation}
where $\Gamma \sim \np \sigma_{\rm int} V$ is the encounter rate, with $\sigma_{\rm int}$ the interaction cross-section. Combining this with the scaling for angular momentum kicks and adopting the gravitational focusing cross section for binary–single encounters  \citep{Heggie1975, HeggieHut2003}
gives:
\begin{equation}\label{eq:De_scaling}
	D_e \propto \frac{(1 - e^2)^2}{e^2} \np \massp^2,
\end{equation}
up to weak dependencies on the velocity dispersion and orbital parameters. This shows that, owing to the quadratic dependence on $\massp$, diffusion is dominated by the most massive perturbers present.  The dependence on eccentricity, on the other hand, reflects the susceptibility of the binary to angular momentum perturbations. As a result, even a sparse population of massive objects can drive significant stochasticity in the eccentricity evolution.

In massive elliptical galaxies, the primary hosts of
the most massive MBHBs targeted by PTAs, the population of potential massive perturbers is dominated by long-lived stellar systems rather than gas structures. These galaxies are typically gas-poor, with little cold molecular material and a correspondingly low abundance of giant molecular clouds, particularly in their central regions \citep[e.g.][]{Young2011, Davis2019}. The dominant population of massive perturbers consists of globular clusters, with typical masses $(10^5-10^6)\msun$ and numbers in the range $10^3\text{--}10^4$. These clusters are long-lived and a fraction of them can be found in the central regions.

In addition to globular clusters, more massive but rarer perturbers are expected in the form of infalling satellites and stripped galactic nuclei. Hierarchical galaxy assembly naturally leads to the presence of such objects, with masses in the range $(10^6\text{--}10^8)\msun$, which can survive tidal stripping and reach the inner kiloparsec
\citep{Pfeffer2014, Antonini2015}. These systems may also host intermediate-mass black holes, further enhancing their dynamical impact. While their number density is low compared to globular clusters, their large masses make them particularly efficient at perturbing binary orbits.

Intermediate-mass black holes (IMBHs), with masses $\sim(10^3\text{--}10^5)\msun$, may also act as perturbers if delivered to the nuclear region by infalling star clusters or satellite remnants. Although their existence and overall abundance is uncertain, their compactness and relatively high masses make them potentially efficient agents of angular momentum perturbations \citep{MillerColbert2004, Greene2020}.

Given the quadratic dependence of the diffusion coefficient on perturber mass, $D_e \propto \np \massp^2$, the stochastic evolution is therefore expected to be dominated by the high-mass tail of the perturber population. This implies that rare, massive encounters, rather than the cumulative effect of numerous low-mass objects, are likely to control the degree of stochasticity in the orbital properties of black hole binaries in massive ellipticals. This dominance is set by the slope of the perturber mass function: for a population with $\mathrm{d}N/\mathrm{d}M \propto M^{-\alpha}$, the mass-weighted diffusion integral $\int M^2\,(\mathrm{d}N/\mathrm{d}M)\,\mathrm{d}M$ receives a contribution per logarithmic mass interval $\propto M^{3-\alpha}$, so it is controlled by the high-mass end for $\alpha < 3$, becomes scale-free at $\alpha \simeq 3$, and is dominated by low-mass objects only for the unrealistically steep case $\alpha > 3$. Realistic perturber populations, with globular cluster and stripped-nucleus mass-function slopes $\alpha \approx 1.5$--$2$ \citep{Forbes2018, Pfeffer2014}, lie in the high-mass-dominated regime, so the assumption that diffusion is governed by the most massive perturbers is robust. The present work isolates this per-mass dependence directly by adopting equal-mass perturbers at fixed total mass.

The total mass in compact stellar systems within the binary's sphere of
influence is expected to be tied to the host galaxy's assembly history. Specifically, the
combined mass in globular clusters, stripped nuclei, and satellite remnants
scales with the host galaxy stellar mass, while the partition of this mass
among individual objects depends on the cluster and satellite mass functions
\citep[e.g.][]{Harris2015,
	Pfeffer2014, Antonini2015}. Varying $\massp$ at fixed
$M_{\rm total} \propto \np \massp$ therefore corresponds to redistributing
a fixed dynamical mass budget across objects of different individual masses,
asking whether it resides in many low-mass globular clusters or a few
massive infalling satellites, rather than changing the total available
perturber reservoir. Under this constraint, the diffusion coefficient scales
as $D_e \propto \massp$, while the encounter rate decreases as
$\Gamma \propto \massp^{-1}$ (this holds in both the gravitational
focusing and geometric cross-section regimes when $\massp \ll \massb$,
since $\Gamma \propto \np$ in both cases). The diffusion approximation
further requires that many encounters occur per binary orbital period,
$\Gamma\, t_{\rm orb} \gg 1$, so that the orbital evolution proceeds as a
continuous stochastic process \citep{PeretsAlexander2008}. At fixed
$M_{\rm total}$, increasing $\massp$ reduces $\Gamma$ until eventually
$\Gamma\, t_{\rm orb} \sim 1$, defining a natural threshold at which
individual encounters become macroscopically significant and the continuous
diffusion description breaks down. This motivates probing both sides of
this transition at fixed total perturber mass, as characterised in
Section~\ref{sec:theory_regimes}.

\subsection{Diffusive and impulsive scattering regimes}\label{sec:theory_regimes}
While the diffusive regime describes the cumulative effect of many distant encounters, this approximation breaks down when single, close passages can produce macroscopic changes in the orbital elements. From Equation~\eqref{eq:angmom_kick}, the angular momentum kick grows as $b$ decreases, but the impulse approximation breaks down for $b \lesssim a$, where the encounter transitions to full binary--single scattering \citep{Heggie1975, HeggieHut2003}. Setting $b \sim a$ therefore gives the maximum per-encounter kick within the impulse approximation. Deep within the influence radius, a perturber falling inward is gravitationally accelerated by the binary. By energy conservation, a perturber reaching $r \sim a$ from a distant starting point acquires a velocity $V(a) \sim \sqrt{2Gm_{b}/a} \sim \velb$ since the escape velocity at $r=a$ is comparable to the binary's own orbital velocity. In this gravitational focusing regime, the perturber velocity is therefore $V \sim \velb$ so the single-encounter kick approaches:
\begin{equation}
	\left(\frac{\Delta J}{J}\right)_{\rm max} \sim \frac{\massp}{\massb} \equiv \mu_{\rm p}.
\end{equation}
The corresponding maximum change in eccentricity from a single event is therefore $\Delta e_{\rm single} \sim [(1-e^2)/e] \mu_{\rm p}$. When $\Delta e_{\rm single}$ is small compared to the run-to-run eccentricity scatter of the $N$-body realisations, individual encounters leave no discernible imprint and the orbital evolution proceeds purely via the cumulative diffusion governed by Equation~\eqref{eq:De_scaling}. However, if the perturber mass is sufficiently large such that $\Delta e_{\rm single}$ approaches or exceeds this variance, a single close encounter can impulsively deliver an observable jump in eccentricity. This physical distinction defines a transition from a continuous diffusive process to a discrete impulsive scattering regime governed by the mass ratio $\mu_{\rm p}$. The threshold is not a function of $\mu_{\rm p}$ alone: the single-encounter eccentricity kick $\Delta e_{\rm single} \sim [(1-e^2)/e]\,\mu_{\rm p}$ also carries an explicit eccentricity dependence through its prefactor. As $e \to 0$ this prefactor diverges, so even a modestly small $\mu_{\rm p}$ can deliver a large jump, and the same perturber population may behave diffusively at high $e$ and impulsively at low $e$. We emphasise that this single-encounter amplitude depends only on the instantaneous orbital state $\{a, \velb, J\}$ of the bound pair, and not on any accumulated post-formation evolution. It therefore quantifies how much a single perturber passage can shift the eccentricity as the binary forms and its angular momentum is set: since the formation eccentricity is fixed by the angular momentum of the pair at $t_{\rm b}$ through $e^2 = 1 - J^2/J_c^2$, the relevant encounters are those occurring up to and around binary formation, rather than post-formation binary--single scatterings of an already-hardened binary. Throughout this paper we diagnose the transition operationally through this single-encounter amplitude relative to the inter-realisation scatter; the rate condition $\Gamma\,t_{\rm orb}\gg1$ introduced above sets the regime of validity of the continuous diffusive description rather than serving as the measured threshold.

\section{Methodology}\label{sec:methodology}
\subsection{\texorpdfstring{\(N\)-body Model}{N-body Model}}
We select a merger from the sample of major mergers identified in the IllustrisTNG100-1 cosmological simulation by \citet{fastidio2024}. Specifically, we adopt their Merger~6, drawn from the TNG100-1 snapshot at redshift $z\approx0.92$ (look-back time $\approx7.6\gyr$), a major merger with a highly eccentric bound orbit and a relatively small apocentre distance of $46.36\,\mathrm{kpc}$; cosmological simulations indicate that such nearly radial encounters are typical of major mergers hosting PTA-relevant MBHBs~\citep{fastidio2024,fastidio2025realistic}, and the short orbital period ensures that the system proceeds to binary formation and hardening within a tractable computational time. Following \citet{fastidio2024}, the progenitors are re-modelled as purely collisionless systems: each galaxy consists of a stellar bulge, a dark matter (DM) halo, and a central MBH, with no gas component. The merger is not uniformly dry: the secondary is gas-poor, but the more massive primary retains a star-forming central reservoir, with a gas fraction $f_{\rm gas}\equiv M_{\rm gas}/(M_{\rm gas}+M_\star)\approx0.19$ within its stellar half-mass radius (mass-weighted $f_{\rm gas}\approx0.15$ for the pair). We nonetheless adopt a collisionless treatment because the most massive PTA-relevant MBHBs are hosted by gas-poor early-type galaxies~\citep[e.g.][]{Young2011, Davis2019}, and because the perturbers we study, long-lived stellar and compact-object systems such as globular clusters and stripped nuclei, are collisionless regardless of the ambient gas. We return to the dynamical role of the neglected gas in Section~\ref{sec:discussion} (caveat~vi). The stellar bulge and DM halo both follow a Hernquist profile \citep{hernquist1990analytical}:
\begin{equation}
	\rho(r) = \frac{M}{2\pi a^3} \frac{a}{r} \frac{1}{(1+r/a)^3},
\end{equation}
where $M$ is the total mass and $a$ is the scale radius. The corresponding model parameters for the primary and secondary galaxies are listed in Table~\ref{tab:params_combined}. The two galaxies are placed at the apocentre of their mutual Keplerian orbit, with an initial orbital eccentricity $e_{\rm orb} = 0.987$.

\begin{table}
	\centering
	\caption{Model parameters of the primary and secondary galaxies: dark matter halo mass $M_{\text{halo}}$, stellar bulge mass $M_{\text{bulge}}$, central MBH mass $M_{\text{bh}}$, and scale radii of the stellar bulge ($a_{\text{bulge}}$) and DM halo ($a_{\text{halo}}$).}
	\label{tab:params_combined}
	\begin{tabular}{lcc}
		\hline
		Parameter                    & Primary                & Secondary              \\
		\hline
		$M_{\text{halo}}$ [$\msun$]  & $1.290 \times 10^{13}$ & $6.928 \times 10^{10}$ \\
		$M_{\text{bulge}}$ [$\msun$] & $3.162 \times 10^{11}$ & $1.056 \times 10^{11}$ \\
		$M_{\text{bh}}$ [$\msun$]    & $1.748 \times 10^{9}$  & $1.347 \times 10^{9}$  \\
		$a_{\text{bulge}}$ [kpc]     & 3.268                  & 1.529                  \\
		$a_{\text{halo}}$ [kpc]      & 59.32                  & 3.243                  \\
		\hline
	\end{tabular}
\end{table}

The smooth bulge and halo components were sampled using a continuous mass-refinement scheme rather than equal-mass particles. In this procedure, the particle mass is assigned as a smooth function of an orbital-radius proxy, taken here to be the circular-orbit radius corresponding to the particle energy, $R_{\rm circ}(E)$. Particles on more tightly bound orbits are assigned lower masses, while particles associated with the outer regions are assigned progressively larger masses, with a smooth transition controlled by a characteristic scale and slope. The overall normalisation is then rescaled so that each component recovers its target Hernquist mass. This refinement increases the particle number and reduces discreteness noise in the central regions relevant for the MBH dynamics, while keeping the total particle number computationally tractable. The resulting median stellar-bulge particle mass is $2.15\times10^{5}\msun$ with a median softening of $25.05\pc$, while the DM halo has a median particle mass of $8.75\times10^{6}\msun$ and a median softening of $159.85\pc$ (full ranges in Table~\ref{tab:particle_inventory_softening}).

\begin{figure}
	\centering
	\includegraphics[width=0.45\textwidth]{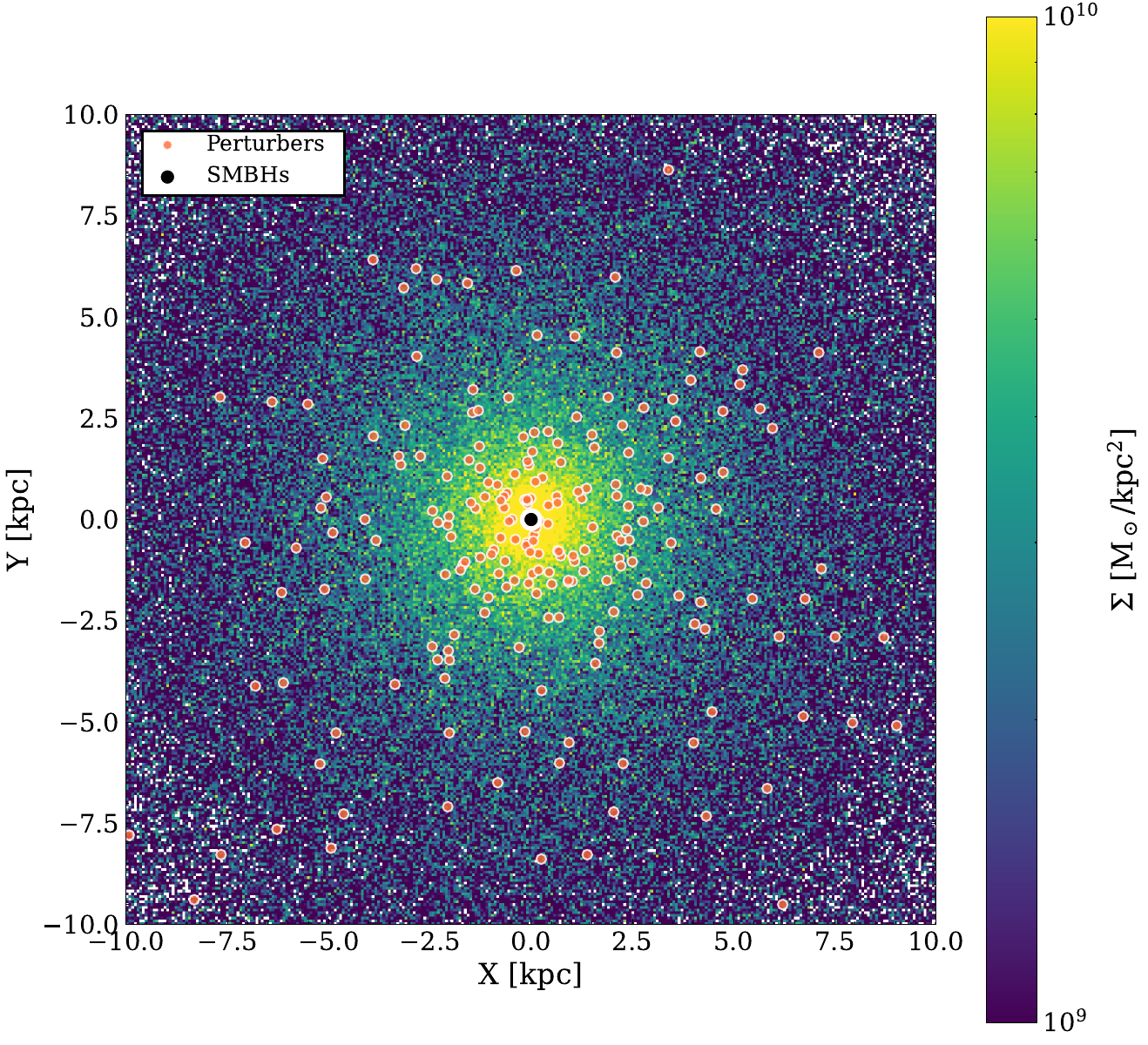}
	\caption{Projected surface density map of the primary bulge in the first realisation of the $10^{8}\msun$ scenario, at the start of the simulation. Orange circles mark the 316 perturbers sampled from the same Hernquist distribution as the bulge; the black circle marks the central MBH. The initial setup for the $10^{7}\msun$ scenario is statistically identical, but with 3162 lower-mass perturbers.}
	\label{fig:initial_setup}
\end{figure}

We follow the evolution of the $N$-body models with the \texttt{Griffin} $N$-body code \citep{dehnen_2014}, which uses a Fast Multipole Method (FMM) scheme that monitors local force errors and adaptively chooses expansion parameters to deliver accuracies comparable to direct summation, while retaining an approximately $\mathcal{O}(N)$ scaling with particle number. We also introduce a population of massive perturbers embedded in the primary galaxy's bulge, sampled from the same Hernquist distribution as the stellar bulge; four independent realisations are generated for each scenario, with the no-perturber control and perturber runs sharing the same parent galaxy realisations and orbital initial conditions. This matched setup ensures that differences between scenarios primarily reflect the added perturber population rather than stochastic differences in the underlying $N$-body sampling at the adopted resolution. Both perturber populations are drawn from the same Hernquist distribution, with no additional central concentration imposed on the more massive set. Fixing the spatial distribution across the two suites ensures that the $10^{7}\msun$ and $10^{8}\msun$ scenarios differ only in perturber mass, so that any difference in the measured scatter can be attributed to the perturber--MBHB mass ratio $\mu_{\rm p}$ rather than to a different radial distribution. In adopting it we neglect the pre-existing mass segregation the perturbers would have acquired over the galaxy's prior assembly history: because dynamical friction acts faster on heavier objects, $t_{\rm DF}\propto\massp^{-1}$, the $10^{8}\msun$ perturbers would in reality have segregated towards the galactic centre more rapidly and reached binary formation more centrally concentrated than the lighter set, a head start that our comparatively short merger run cannot build up in situ. This omission is conservative for the $10^{8}\msun$ case, in the sense that a self-consistent treatment would bring more massive perturbers within the binary's sphere of influence at $t_{\rm b}$, plausibly strengthening rather than weakening the measured effect.
We assign a fixed target mass fraction $f_{\rm target}$ of the primary bulge mass $M_{\rm bulge,primary}$ to the perturber population, so that the total perturber mass is:
\begin{equation}
	M_{\rm total} = f_{\rm target} M_{\rm bulge,primary}.
\end{equation}
To conserve the total bulge mass, we remove an equivalent mass $M_{\rm total}$ in stellar bulge particles from the primary, so that the perturbers replace a fraction $f_{\rm target}$ of the original smooth bulge component. Assuming equal-mass perturbers of mass $\massp$, the corresponding number is:
\begin{equation}
	N = \frac{f_{\rm target}\, M_{\rm bulge,primary}}{\massp}.
\end{equation}
For our fiducial choice $M_{\rm bulge,primary} = 3.162 \times 10^{11}\msun$ and $f_{\rm target} = 0.1$, the resulting values of $N$ for different perturber masses are listed in Table~\ref{tab:perturber_numbers}. We adopt $f_{\rm target}=0.1$ as a generous upper estimate, chosen to maximise the perturber-driven dynamical effect at fixed $\massp$. The realised per-component particle counts, masses, and softening lengths for a representative realisation are summarised in Table~\ref{tab:particle_inventory_softening}, and the initial spatial distribution of the $10^{8}\msun$ perturber population is shown in Fig.~\ref{fig:initial_setup}.

\begin{table}
	\centering
	\caption{Number of equal-mass perturbers for different perturber masses, assuming a primary bulge mass
		$M_{\rm bulge,primary} = 3.162 \times 10^{11}\msun$ and a target fraction $f_{\rm target} = 0.1$
		(i.e. $M_{\rm total} = 3.162 \times 10^{10}\msun$). Also listed are the corresponding
		binary mass ratios $\mu_{\rm p} \equiv \massp/M_{\rm b}$ for $M_{\rm b} = M_{\rm BH,1}+M_{\rm BH,2} = 3.095\times10^{9}\msun$.}
	\label{tab:perturber_numbers}
	\begin{tabular}{lccc}
		\toprule
		$\massp$ [$\msun$] & $M_{\rm total}$ [$\msun$] & $N = M_{\rm total} / \massp$ & $\mu_{\rm p}$        \\
		\midrule
		$10^{8}$           & $3.162 \times 10^{10}$    & 316                          & $3.2 \times 10^{-2}$ \\
		$10^{7}$           & $3.162 \times 10^{10}$    & 3,162                        & $3.2 \times 10^{-3}$ \\
		\bottomrule
	\end{tabular}
\end{table}

% \begin{table}
% \centering
% \caption{Number of equal-mass perturbers for different perturber masses, assuming a primary bulge mass
% $M_{\rm bulge,primary} = 3.162 \times 10^{11}\msun$ and a target fraction $f_{\rm target} = 0.1$
% (i.e. $M_{\rm total} = 3.162 \times 10^{10}\msun$).}
% \label{tab:perturber_numbers}
% \begin{tabular}{lcc}
% \toprule
% $\massp$ [$\msun$] & $M_{\rm total}$ [$\msun$] & $N = M_{\rm total} / \massp$ \\
% \midrule
% $10^{8}$ & $3.162 \times 10^{10}$ & 316   \\
% $10^{7}$ & $3.162 \times 10^{10}$ & 3,162 \\
% \bottomrule
% \end{tabular}
% \end{table}

\begin{table*}
	\centering
	\caption{Per-component particle number, masses, and softening lengths for a sample realisation of each scenario. Masses and softenings are in physical units, using one code mass unit $=3.162\times10^{11}\msun$ and one code length unit $=1\,\mathrm{kpc}$. For mass-refined standard particles, the tabulated value is the median, with the full range in parentheses. Perturbers are represented as softened extended particles rather than unresolved point masses, so the $10^{8}\msun$ case should be interpreted as a conservative test of near-impulsive massive substructure at the adopted force resolution.}
	\label{tab:particle_inventory_softening}
	\begin{tabular}{llrcc}
		\toprule
		Scenario & Component & $N$ & Particle mass [$\msun$] & Softening [pc] \\
		\midrule
		\multirow{3}{*}{No perturbers}
		& MBHs  & 2         & $(1.35\text{--}1.75)\times10^{9}$ & $2.63\text{--}3.00$ \\
		& Bulge & 2,000,949 & $2.15\times10^{5}$ $(3.41\times10^{4}\text{--}3.43\times10^{5})$ & $25.05$ $(9.98\text{--}31.64)$ \\
		& Halo  & 1,508,056 & $8.75\times10^{6}$ $(1.34\times10^{6}\text{--}1.40\times10^{7})$ & $159.85$ $(62.57\text{--}202.28)$ \\
		\midrule
		\multirow{4}{*}{$10^{7}\msun$ perturbers}
		& MBHs       & 2         & $(1.35\text{--}1.75)\times10^{9}$ & $2.63\text{--}3.00$ \\
		& Bulge      & 1,850,955 & $2.15\times10^{5}$ $(3.41\times10^{4}\text{--}3.43\times10^{5})$ & $25.05$ $(9.98\text{--}31.64)$ \\
		& Halo       & 1,508,056 & $8.75\times10^{6}$ $(1.34\times10^{6}\text{--}1.40\times10^{7})$ & $159.85$ $(62.57\text{--}202.28)$ \\
		& Perturbers & 3,162     & $1.00\times10^{7}$ & $85.47$ \\
		\midrule
		\multirow{4}{*}{$10^{8}\msun$ perturbers}
		& MBHs       & 2         & $(1.35\text{--}1.75)\times10^{9}$ & $2.63\text{--}3.00$ \\
		& Bulge      & 1,851,050 & $2.15\times10^{5}$ $(3.41\times10^{4}\text{--}3.43\times10^{5})$ & $25.05$ $(9.98\text{--}31.64)$ \\
		& Halo       & 1,508,056 & $8.75\times10^{6}$ $(1.34\times10^{6}\text{--}1.40\times10^{7})$ & $159.85$ $(62.57\text{--}202.28)$ \\
		& Perturbers & 316       & $1.00\times10^{8}$ & $266.93$ \\
		\bottomrule
	\end{tabular}
\end{table*}

In \texttt{Griffin}, all standard particles (both stellar and DM) are assigned individual softenings according to:
\begin{equation}
	\varepsilon = \varepsilon_{\rm ref} \left(\frac{m}{\tilde m_\ast}\right)^{1/2},
\end{equation}
where $\varepsilon_{\rm ref}=3 \,\mathrm{pc}$, and $\tilde m_\ast$ is the global median particle mass of the combined bulge and DM halo particles of the primary and secondary galaxies. The softenings listed in Table~\ref{tab:particle_inventory_softening} are the values assigned in the matching input snapshots. The exponent $1/2$ ensures that the maximum pairwise force $Gm/\varepsilon^{2}=G\tilde{m}_{\ast}/\varepsilon_{\rm ref}^{2}$ is independent of particle mass, preventing massive particles from dominating over the mean field \citep{dehnen_2014, power_2003}. Separately, the perturbers are given softenings via:
\begin{equation}
	\varepsilon = \mathcal{C}\,\varepsilon_{\rm ref} \left(\frac{m}{\tilde m_\ast}\right)^{1/2},
\end{equation}
with an empirical compactness factor $\mathcal{C}=0.5$. These particles represent a generic population of massive perturbers in the galactic nucleus, such as dense stellar clusters~\citep{perets2006massive, Perets2007, PeretsAlexander2008} or intermediate-mass black holes (IMBHs) embedded in nuclear star clusters (NSCs)~\citep{mastrobuono2014effects}, which can dominate relaxation and loss-cone refilling near MBHs. Giant molecular clouds of comparable mass could play an analogous role in gas-rich mergers, including the star-forming primary of the present system (Section~\ref{sec:methodology}); they are not captured by our collisionless model, and we discuss the consequences in Section~\ref{sec:discussion} (caveat~vi). The two perturber masses are selected to probe either side of the transition between diffusive and near-impulsive scattering, as characterised by the binary mass ratio $\mu_{\rm p} \equiv \massp/M_{\rm b}$ (Table~\ref{tab:perturber_numbers}). At $\mu_{\rm p} \approx 3.2\times10^{-3}$, the $10^{7}\msun$ case is expected to lie in the diffusive limit, where individual encounters contribute small velocity kicks that accumulate gradually. At $\mu_{\rm p} \approx 3.2\times10^{-2}$, the $10^{8}\msun$ case approaches the near-impulsive regime, where individual close encounters may deliver kicks $\Delta v/v \sim \mu_{\rm p}$ (from the impulse approximation) large enough to appreciably alter the BH trajectory on a single passage, subject to the finite perturber softening scale listed in Table~\ref{tab:particle_inventory_softening}.

\subsection{Binary Formation Criterion}
We track the specific two-body orbital energy
\begin{equation}
	\mathcal{E} = \frac{1}{2}|\dot{\boldsymbol{r}}|^{2} - \frac{G \massb}{r},
\end{equation}
where $\boldsymbol{r}$ is the relative separation of the MBH pair and
$\massb = M_{\rm BH,1} + M_{\rm BH,2}$.
The pair is gravitationally bound when $\mathcal{E} < 0$ and $e \in (0,1)$.
Along the nearly radial merger orbit, the MBHs undergo repeated close passages
before cumulative stellar-dynamical friction drives the apocentre inside the
influence sphere and permanent binding is established.

Following the initial chaotic binding phase~\citep{gualandris2022,fastidio2024, gualandris2026converging},
the instantaneous eccentricity continues to oscillate before settling.
We therefore define the binary formation time $t_{\rm b}$ as the first local
minimum of $e(t)$ following permanent binding~\citep[cf.][]{fastidio2024},
which marks the onset of the classical hardening phase~\citep{begelman_1980,quinlan1996}.
The formation eccentricity is the instantaneous value $e_{\rm b}\equiv e(t_{\rm b})$,
and the scatter quoted throughout, $\sigma_e\equiv\mathrm{SD}(\{e_{{\rm b},i}\})$, is
its standard deviation across the four realisations $i$ of a given scenario: a measure
of inter-realisation scatter at binary formation, not a time-averaged or
post-formation quantity.

\section{Results}\label{sec:results}

Before comparing the perturber scenarios, we first characterise the level of run-to-run scatter inherent to the $N$-body setup in the absence of massive perturbers (we use $t_{\rm b}$ for the binary formation time defined in Section~\ref{sec:methodology}, and $e_{\rm b} \equiv e(t_{\rm b})$ for the formation eccentricity throughout). The eccentricity standard deviation for the no-perturber control case yields $\sigma_e \approx 0.11$ at a median formation eccentricity $\langle e_{\rm b}\rangle \approx 0.64$ (Fig.~\ref{fig:formation_summary}). This value is consistent, in normalisation, with the empirical Poisson noise floor expected at this resolution from the $\sigma_e \propto N^{-1/2}$ scaling measured by \citet{gualandris2026converging}. Because the present control suite uses a single resolution, we do not measure the scaling directly. The comparison is approximate, since \citet{gualandris2026converging} employ single-component stellar-bulge models whereas our setup is multi-component. We define $\sigma_e \approx 0.11$ as the numerical noise floor: the reference level of eccentricity scatter against which perturber-driven stochasticity can be assessed.

\subsection{Orbital evolution under massive perturbers}\label{sec:results_orbital}

\begin{figure*}
	\centering
	\includegraphics[width=0.85\textwidth]{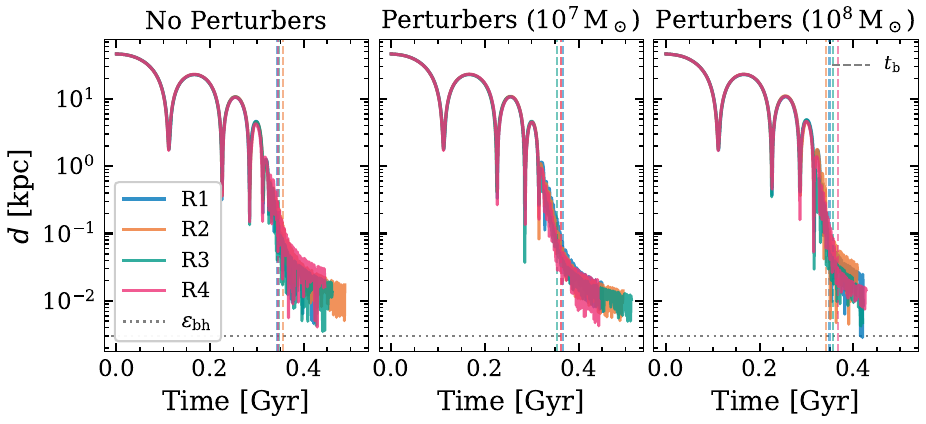}
	\caption{MBH separation $d$ as a function of time for all four realisations in each scenario: no perturbers (left), $10^7\msun$ perturbers (centre), $10^8\msun$ perturbers (right). Individual realisations are shown as thin coloured lines, and vertical dashed lines mark the binary formation time $t_{\rm b}$ of each realisation. The dotted horizontal line indicates the MBH softening $\varepsilon_{\rm bh}$.}
	\label{fig:bh_distance_comparison}
\end{figure*}

\begin{figure*}
	\centering
	\includegraphics[width=0.85\textwidth]{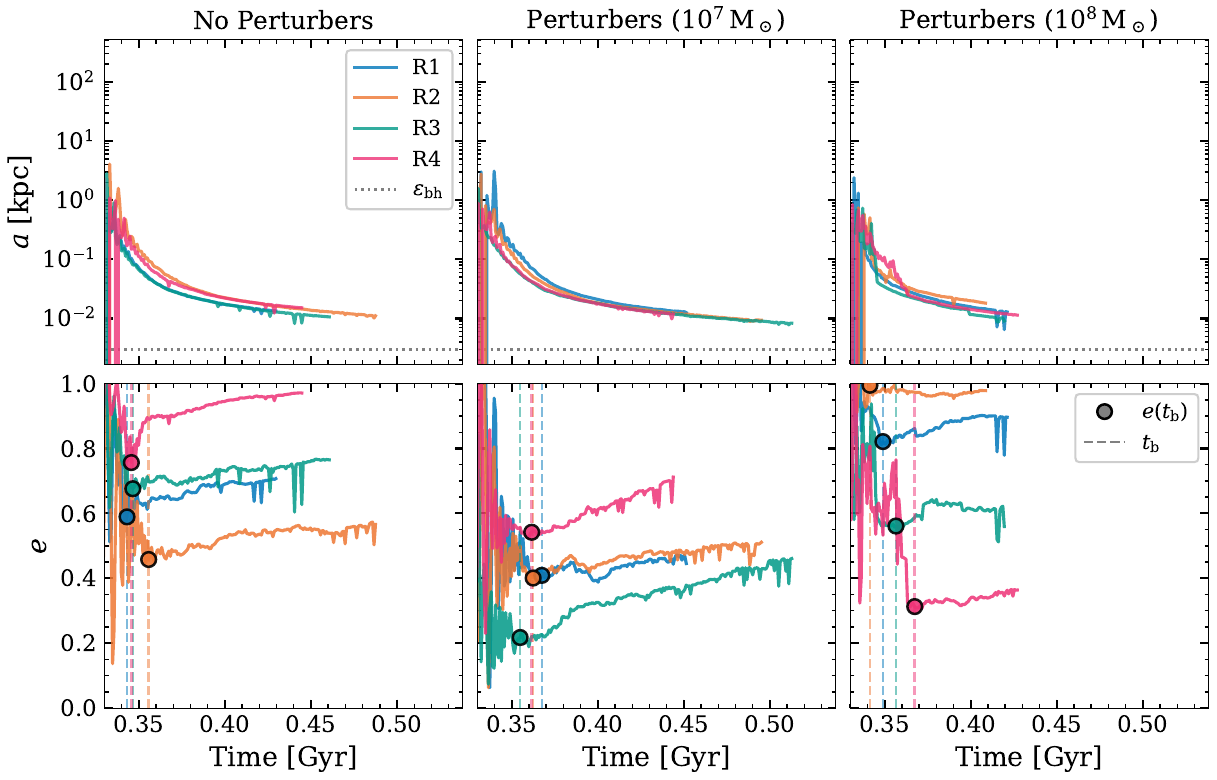}
	\caption{Evolution of the Keplerian semi-major axis $a$ (top row) and orbital eccentricity $e$ (bottom row) for all four realisations in each scenario: no perturbers (left), $10^7\msun$ perturbers (centre), $10^8\msun$ perturbers (right). Vertical dashed lines mark the time at binary formation $t_{\rm b}$ for each realisation. The no-perturber and $10^7\msun$ cases display smooth, coherent evolution across realisations, whereas the $10^8\msun$ case exhibits discrete jumps in $a$ and a wider spread in $e(t_{\rm b})$.}
	\label{fig:orb_elem_comparison}
\end{figure*}

The separation between the MBHs and the two-body Keplerian orbital elements
are shown in Fig.~\ref{fig:bh_distance_comparison} and
Fig.~\ref{fig:orb_elem_comparison} respectively. The MBH separation evolves
similarly across all three scenarios, with the large-scale inspiral governed
by the smooth stellar background rather than the perturber population; the
Keplerian elements are required to resolve the perturber-driven differences.

The semi-major axis and eccentricity evolution for the no-perturber and $10^{7}\msun$ cases are smooth and comparable across realisations, as expected for the diffusive scattering regime ($\mu_{\rm p} \approx 3.2\times10^{-3}$) in which individual encounters contribute velocity kicks $\Delta v/v \sim \mu_{\rm p}$ too small to leave a discernible imprint on the inspiral trajectory. The $10^{7}\msun$ case tends towards lower eccentricities at formation; whether this offset is physical or attributable to numerical stochasticity is examined in Section~\ref{sec:results_phys}.

The $10^{8}\msun$ scenario exhibits noticeable differences. Two of the four realisations display discrete jumps in $a(t)$, accompanied by corresponding offsets in the orbital energy (Fig.~\ref{fig:aligned_dynamics}), the signature of individual close perturber--BH encounters in the near-impulsive regime. All four realisations show a wider spread in eccentricity relative to the other scenarios, with the same two realisations experiencing the most obvious eccentricity jumps.
We will further quantify this scatter relative to the numerical noise floor in
Section~\ref{sec:results_phys}.

\subsection{Physical stochasticity at binary formation}\label{sec:results_phys}

\begin{figure*}
	\centering
	\includegraphics[width=0.85\textwidth]{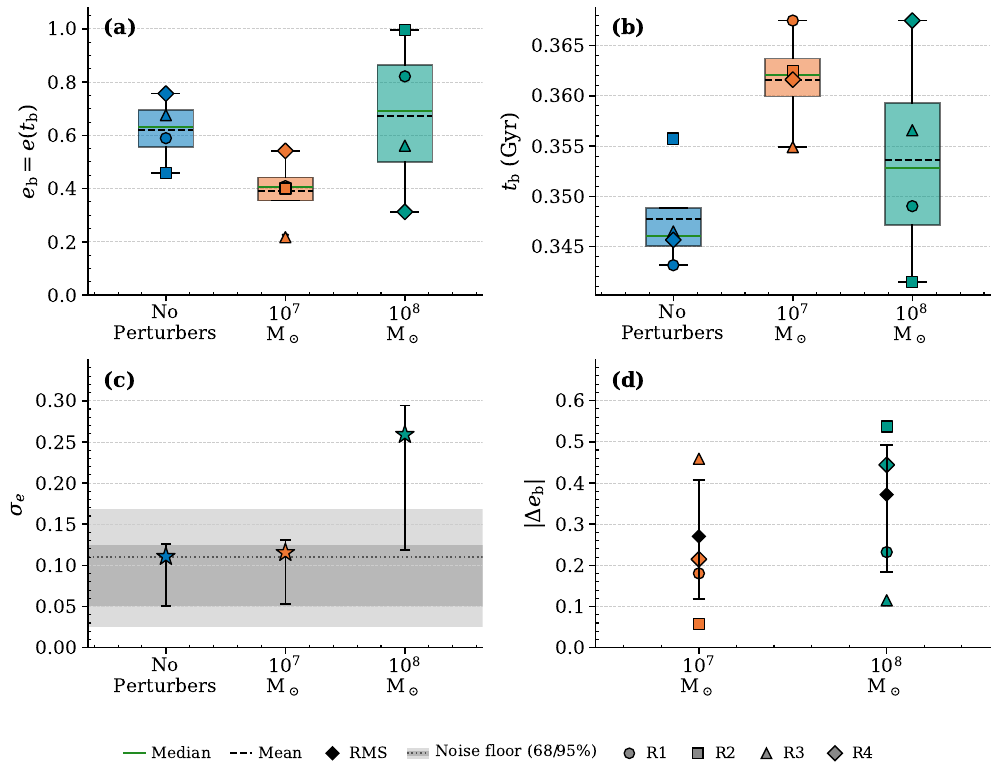}
	\caption{Formation properties across perturber scenarios. In panels (a), (b) and (d), coloured markers show the four realisations (R1--R4: circle, square, triangle, diamond), coloured by scenario; in (a) and (b) the box spans the interquartile range, with the solid green line the median and the dashed line the mean. (a) Formation eccentricity $e_{\rm b}=e(t_{\rm b})$. (b) Binary formation time $t_{\rm b}$. (c) Eccentricity scatter $\sigma_e$ (star per scenario). The grey band is the sampling distribution of $\sigma_e$ under the numerical-noise-floor null, with the four eccentricities drawn from a Gaussian of scatter $\sigma_{\rm true}=0.11$; nested shades give the central $68$ and $95$ per cent intervals (scaled-$\chi$, three degrees of freedom), centred on the floor (dotted line). Error bars repeat the $68$ per cent interval anchored on each measured $\sigma_e$. The control and $10^{7}\msun$ values fall within the band; the $10^{8}\msun$ value lies above it. (d) Absolute paired shift $|\Delta e_{\rm b}|=|e_{\rm b}^{\rm pert}-e_{\rm b}^{\rm ctrl}|$ relative to the matched control, with the black diamond the RMS shift and its $95$ per cent bootstrap interval.}
	\label{fig:formation_summary}
\end{figure*}

The $10^{8}\msun$ perturber scenario is the only case in which the
eccentricity scatter at binary formation measurably exceeds the noise floor. Eccentricities and time at binary formation, eccentricity
scatter, and matched formation-eccentricity shifts are summarised in
Fig.~\ref{fig:formation_summary}, with the measured $\sigma_e$ values also
listed in Table~\ref{tab:regime_summary}. Because the perturber and
no-perturber runs are matched realisation by realisation, the paired
quantity $|\Delta e_{\rm b}|=|e_{\rm b}^{\rm pert}-e_{\rm b}^{\rm ctrl}|$
removes much of the control run-to-run variation and isolates the amplitude
of the perturber-induced change in the eccentricity at binary formation for
the same realisation.

For the $10^{7}\msun$ case, $\sigma_e \approx 0.115$ is
indistinguishable from the control ($\sigma_e \approx 0.11$), with no
measurable increase in stochasticity in eccentricity. The median
$e(t_{\mathrm{b}}) \approx 0.41$ is lower than the control ($\approx 0.64$),           
though with only four realisations this offset should be interpreted
cautiously; we discuss its potential origin in Section~\ref{sec:discussion}.
The median $t_{\mathrm{b}} \approx 0.362\,\mathrm{Gyr}$ is marginally later
than the control ($\approx 0.346\,\mathrm{Gyr}$) with little run-to-run
scatter. For the $10^{8}\msun$ scenario, $e(t_{\mathrm{b}})$ spans
$\approx 0.30$--$0.99$ with $\sigma_e \approx 0.26$, exceeding the control
by a factor of $\approx 2.4$, and $t_{\mathrm{b}}$ ranges over
$\approx 0.341$--$0.368\,\mathrm{Gyr}$ across different realisations.

The paired eccentricity shifts show that the $10^{7}\msun$ perturbers mainly produce
modest changes relative to the matched controls, whereas the
$10^{8}\msun$ case produces systematically larger realisation-dependent
offsets (Fig.~\ref{fig:formation_summary}d). This indicates enhanced
stochasticity about the matched no-perturber control.

\begin{table}
	\centering
	\caption{Predicted single-encounter eccentricity kicks $\Delta e_{\rm single}
			\sim [(1-e^2)/e]\,\mu_{\rm p}$ evaluated at the control median eccentricity
		$\langle e(t_{\rm b})\rangle \approx 0.64$, compared against the measured
		$\sigma_e$ for each scenario.}
	\label{tab:regime_summary}
	\begin{tabular}{lcccl}
		\toprule
		Scenario      & $\mu_{\rm p}$      & $\Delta e_{\rm single}$ & $\sigma_e$ & Regime         \\
		\midrule
		No perturbers & $\cdots$           & $\cdots$                & $0.110$    & Control        \\
		$10^{7}\msun$ & $3.2\times10^{-3}$ & $\sim0.003$             & $0.115$    & Diffusive      \\
		$10^{8}\msun$ & $3.2\times10^{-2}$ & $\sim0.030$             & $0.260$    & Near-impulsive \\
		\bottomrule
	\end{tabular}
\end{table}

The two scenarios probe distinct dynamical regimes, and the appropriate
comparison with $\sigma_e$ differs between them. In the diffusive regime,
no single encounter dominates; instead, the relevant quantity is the
cumulative eccentricity variance accumulated over the inspiral timescale
$\Delta t$. At fixed total perturber mass, $D_e \propto \massp$ (since
$\np \propto \massp^{-1}$; Section~\ref{sec:theory}), so $\sqrt{D_e\,\Delta t}
	\propto \massp^{1/2}$. A rough estimate for the $10^{7}\msun$
case gives $\sqrt{D_e\,\Delta t}\sim 0.03\text{--}0.1$, using
$\Delta e_{\rm single}\sim 0.003$ and $\Gamma\,\Delta t \sim 10^{2}\text{--}10^{3}$
encounters over the $\sim 0.3\,\mathrm{Gyr}$ pre-formation interval;
this sits roughly at or below the noise floor, in line with the observed $\sigma_e \approx 0.115 \approx
	\sigma_{e,{\rm control}}$. In the near-impulsive regime, the eccentricity
evolution is instead dominated by rare, close encounters. Where such an encounter
occurs shortly before $t_{\rm b}$, the kick $\Delta e_{\rm single} \sim
	[(1-e^2)/e]\,\mu_{\rm p}$ is imprinted at binary formation before subsequent
orbital evolution can damp it, making $\Delta e_{\rm single}$ the relevant
estimate of the inter-realisation scatter such an encounter would produce. This interpretation matches the $10^{8}\msun$ case, where the discrete eccentricity jumps in Fig.~\ref{fig:orb_elem_comparison} occur close to $t_{\rm b}$. The single-encounter prediction $\Delta e_{\rm single} \sim 0.030$, evaluated at the control median $e \approx 0.64$, falls short of the perturber-induced scatter $\sqrt{\sigma_e^2 - \sigma_{e,{\rm control}}^2} \approx 0.24$ (treating the perturber-induced and baseline contributions as independent, so that their variances add and the control variance can be subtracted in quadrature) by a factor of ${\sim}8$. Two effects can bridge this gap: the eccentricity-dependent prefactor $(1-e^2)/e$ is a factor of ${\sim}3$ larger at $e \approx 0.30$ (the lowest measured $e_{\rm b}$ in the $10^8\msun$ scenario) than at the control median, so encounters occurring when the binary is near-circular deliver larger per-event kicks; and a few (${\sim}2$--$3$) close passages per realisation, each contributing $\Delta e \sim 0.03$--$0.09$, together bring the expected spread to within a factor of a few of the inferred value, i.e.\ into order-of-magnitude agreement. The $10^8\msun$ runs therefore sit in an intermediate, few-body regime approaching the impulsive limit rather than in the single-kick limit itself. With only four realisations per scenario we do not interpret this enhancement from $\sigma_e$ alone, but from its agreement with the distinct dynamical diagnostics presented in Sections~\ref{sec:results_dynamics} and~\ref{sec:results_brownian} (event-aligned residuals in orbital energy, angular momentum, and torque, and the enhanced centre-of-mass wander), all weaker or absent in the control and $10^{7}\msun$ runs. We assess the statistical significance of the scatter and the sufficiency of the present suites in Section~\ref{sec:discussion} (caveat~i).

\subsection{Dynamical signatures around binary formation}\label{sec:results_dynamics}
\begin{figure*}
	\centering
	\includegraphics[width=0.9\textwidth]{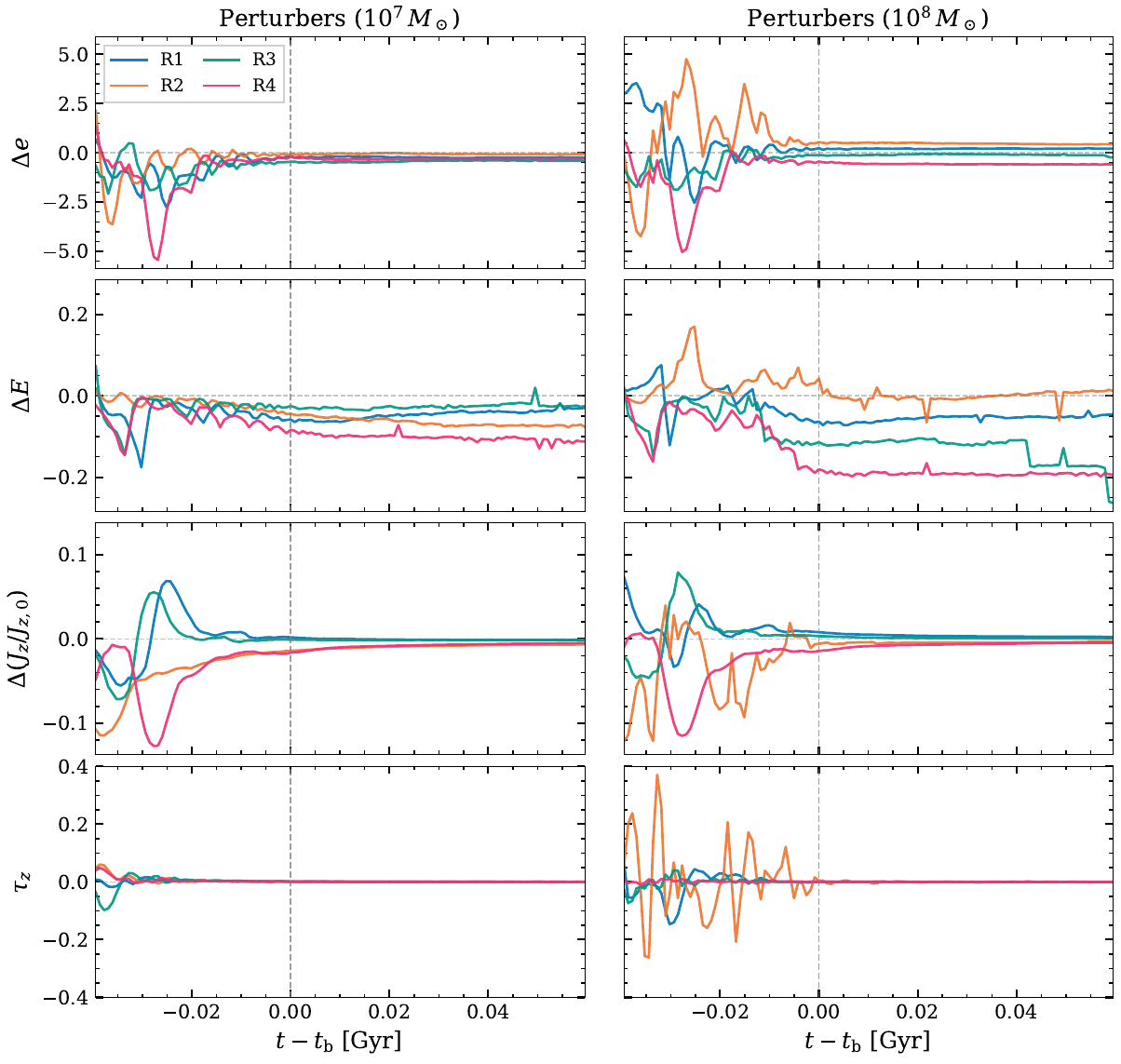}
	\caption{Event-aligned dynamical residuals for the perturbed MBHB runs. Columns show the $10^{7}\msun$ and $10^{8}\msun$ perturber scenarios, and colours denote realisations R1--R4. The horizontal axis is time relative to each realisation's binary formation time, $t-t_{\rm b}$. The four rows show matched residuals relative to the corresponding no-perturber control at the same relative evolutionary phase: eccentricity $\Delta e$, specific orbital energy $\Delta E$, normalised angular-momentum component $\Delta(J_z/J_{z,0})$ and torque component $\tau_z$. The $10^{7}\msun$ case shows relatively smooth residual evolution and weak torques near formation, whereas the $10^{8}\msun$ case exhibits larger realisation-dependent eccentricity offsets, stronger energy and angular-momentum residuals, and larger torque spikes just before $t_{\rm b}$.}
	\label{fig:aligned_dynamics}
\end{figure*}

Fig.~\ref{fig:aligned_dynamics} compares the orbital response of each perturbed
realisation to its matched no-perturber control, after aligning both runs by
their own binary formation times. For each quantity
$Q \in \{e,E,J_z/J_{z,0}, \tau_{z}\}$ we define the relative time:
\begin{equation*}
	t_{\rm rel} \equiv t-t_{\rm b},
\end{equation*}
so that $t_{\rm rel}=0$ corresponds to binary formation in each run. The four
rows of Fig.~\ref{fig:aligned_dynamics} then show the residual between the
perturbed run and its matched control at the same $t_{\rm rel}$:
\begin{equation*}
	\Delta Q(t_{\rm rel}) \equiv Q_{\rm pert}(t_{\rm rel}) - Q_{\rm ctrl}(t_{\rm rel}),
\end{equation*}
where each quantity is evaluated after shifting the corresponding run so that
$t=0$ coincides with its own $t_{\rm b}$. This matched comparison removes much
of the coherent orbital evolution shared by each paired realisation, leaving
the residual response to the perturber population.

The $10^{7}\msun$ realisations show no signs of strong perturbation near binary formation. Although $\Delta e$ can differ substantially
at earlier times, the residual energy evolves relatively smoothly,
while $\Delta(J_z/J_{z,0})$ decays toward small values by $t_{\rm b}$. The
corresponding $\tau_{z}$ are of low amplitudes, in which they are rapidly damped. This behaviour seems to support the formation-eccentricity statistics in Fig.~\ref{fig:formation_summary}: the $10^{7}\msun$ perturbers shift the detailed orbital phase of individual runs but do not produce a measurable increase in stochasticity above the no-perturber noise floor.

In the $10^{8}\msun$ case, the eccentricity residuals remain strongly
realisation dependent through binary formation, and the same interval shows
larger offsets and sharper structure in both $\Delta E$ and
$\Delta(J_z/J_{z,0})$. $\tau_{z}$ also contains large amplitude spikes that
are absent, or much weaker, in the $10^{7}\msun$ case. The enhanced scatter in
$e_{\rm b}$ is therefore accompanied by changes in the binary's energy and
angular-momentum evolution, rather than by eccentricity offsets alone.

The hierarchy between the energy and angular-momentum response is consistent with the scaling arguments of Section~\ref{sec:theory}. The angular-momentum kick (Equation~\ref{eq:angmom_kick}) is only linearly suppressed with impact parameter, $\propto a/b$, so a wider range of passages can contribute to angular-momentum fluctuations. Energy changes from distant encounters are more strongly suppressed, while appreciable changes in the binary binding energy are expected mainly from close passages with $b\sim a$. The larger event-aligned residuals and torque spikes in the $10^{8}\msun$ runs are therefore consistent with a transition toward near-impulsive behaviour, whereas the $10^{7}\msun$ runs remain consistent with diffusive perturbations at the resolution and sample size considered here.

\subsection{Wander of the binary centre of mass}\label{sec:results_brownian}
\begin{figure*}
	\centering
	\includegraphics[width=0.95\textwidth]{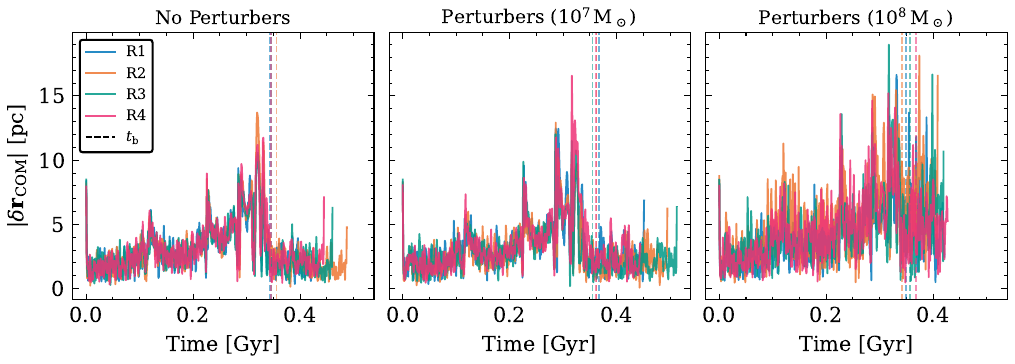}
	\caption{High-frequency wander of the binary centre of mass,
		$|\delta\boldsymbol{r}_{\rm COM}|$, for all four realisations (R1--R4) in
		each scenario: no perturbers (left), $10^{7}\msun$ perturbers (centre),
		$10^{8}\msun$ perturbers (right). The wander is the residual of the
		mass-weighted MBH centre of mass about a Gaussian-smoothed ($1\,\myr$)
		trajectory, isolating small-scale displacements from the common
		kiloparsec-scale merger plunge. Vertical dashed lines mark the binary
		formation time $t_{\rm b}$ of each realisation. The $10^{8}\msun$ case shows
		a systematically elevated wander throughout the pre-formation inspiral, with
		a pooled RMS amplitude a factor of $\approx 1.25$--$1.4$ above the control (window-dependent), whereas the
		$10^{7}\msun$ case is indistinguishable from the no-perturber control.}
	\label{fig:com_wander}
\end{figure*}

The event-aligned residuals of Section~\ref{sec:results_dynamics} track the
binary's internal orbital response to the perturbers. As a complementary diagnostic, we also examined the wander of the binary centre of mass
(COM). We reconstruct the COM trajectory from the BH data as the mass-weighted
mean of the two MBH position vectors,
$\boldsymbol{r}_{\rm COM}=(M_{\rm BH,1}\boldsymbol{r}_1+M_{\rm BH,2}\boldsymbol{r}_2)/\massb$,
and isolate its high-frequency wander $\delta\boldsymbol{r}_{\rm COM}$ as the
residual about a Gaussian-smoothed ($1\,\myr$) trajectory, which subtracts the
kiloparsec-scale merger plunge common to all the runs (Fig.~\ref{fig:com_wander}).
We want to highlight that $\delta\boldsymbol{r}_{\rm COM}$ is not a direct measure of the
pericentre separation, sensitivity to which, through the closest-approach
deflection angle, drives the eccentricity randomisation in the
\citet{rawlings2023} picture. That randomisation is governed by the relative
coordinate $\boldsymbol{r}_1-\boldsymbol{r}_2$, whereas the COM is the
mass-weighted mean position: for our near-equal-mass pair
($M_{\rm BH,2}/M_{\rm BH,1}\approx0.77$) it reduces to
$\delta\boldsymbol{r}_{\rm COM}\approx(\delta\boldsymbol{r}_1+\delta\boldsymbol{r}_2)/2$,
so it captures the common-mode recoil of the barycentre rather than the
differential response of the separation itself. We therefore use it as a
tracer of the parsec-scale perturbation environment: the same close passages
that recoil the barycentre must also perturb the relative orbit, so an increased COM
wander supports the internal-response signatures of
Section~\ref{sec:results_dynamics}.

In the no-perturber control and the $10^{7}\msun$ suite the COM follows a
small-amplitude random walk. Measured over the pre-formation inspiral
($t<t_{\rm b}$), both have a pooled root-mean-square (RMS) wander of
$\approx 4.0\,\pc$ (median $\approx 3.0\,\pc$), with peaks of $\approx 14$--$17\,\pc$
near binary formation. The $10^{8}\msun$ suite shows a modestly higher wander,
with a pre-formation RMS of $\approx 5.0\,\pc$, a factor of $\approx 1.25$ above the
control-level floor (median $\approx 3.8\,\pc$, peaks $\approx 19\,\pc$); the contrast rises to a
factor of $\approx 1.3$ in a $\pm50\,\myr$ window around $t_{\rm b}$ ($\approx 7.1$
versus $\approx 5.4\,\pc$) and to $\approx 1.4$ over the full track. The
$10^{7}\msun$ wander is indistinguishable from the control in every window. The
excess is not confined to the immediate vicinity of $t_{\rm b}$ but persists across
the inspiral, indicating repeated displacements of the binary rather than a single
late kick. The absolute RMS amplitude scales with the $1\,\myr$ smoothing cut and is
pooled over correlated time samples within only four realisations, so we report the
$10^{8}/$control ratio at fixed smoothing: the enhancement persists for smoothing
windows between $0.5$ and $2\,\myr$, and at the fiducial $1\,\myr$ smoothing the
(window-dependent) ratio is $\approx 1.25$--$1.4$. We treat it as qualitative
corroboration of the near-impulsive interpretation rather than a precise ensemble
measurement. Its physical origin is the same as that of the
enhanced $\sigma_e$ (Section~\ref{sec:results_phys}) and torque spikes
(Section~\ref{sec:results_dynamics}): in the near-impulsive regime, close
perturber passages deliver kicks $\Delta v/v\sim\mu_{\rm p}$ that both recoil the
binary barycentre and perturb its internal orbit.

The measured wander combines numerical Brownian motion of the massive bodies
against the finite-mass stellar background with physical kicks from the
perturbers. For the fixed stellar background common to all three suites, the numerical
component depends only on the background particle mass relative to the binary
mass \citep{merritt2001brownian, merritt2007brownian},
which is identical across all three suites by construction
(Table~\ref{tab:particle_inventory_softening}); it is therefore common to every
scenario, and the $\approx 4.0\,\pc$ wander shared by the control and
$10^{7}\msun$ suites measures it directly. At our resolution
($N_{\rm bulge}\approx 2\times10^{6}$, $\tilde m_\ast/\massb\sim 7\times10^{-5}$)
this floor is also expected to be dynamically sub-dominant, consistent with
\citet{bortolas2016brownian}, who find that Brownian motion no longer
significantly affects MBHB evolution once $N\gtrsim10^{6}$. The excess in the
$10^{8}\msun$ case therefore reflects physical perturber kicks rather than
numerical noise.

\section{Discussion}\label{sec:discussion}
The results from Section~\ref{sec:results} are consistent with a transition from diffusive to near-impulsive scattering governed by the perturber--MBHB mass ratio $\mu_{\rm p}$. Only the $10^8 \msun$ scenario, which approaches the near-impulsive regime, shows an enhancement in eccentricity stochasticity above the numerical noise floor. Our main astrophysical conclusion, however, follows from where realistic perturbers fall in $\mu_{\rm p}$ rather than from the statistical strength of this single enhancement (caveat~i).

We measure three eccentricity scatter values ($\sigma_e \approx 0.11$, $0.115$, and $0.26$ for the three scenarios) alongside the dynamical signatures of Section~\ref{sec:results_dynamics}. The near-impulsive interpretation of the $10^8\msun$ case rests on this combination: enhanced $\sigma_e$, event-aligned residuals in $E$, $J_z/J_{z,0}$, and $\tau_z$, and the accompanying $\approx 1.3$ (window-dependent) enhancement in the wander of the binary centre of mass (Section~\ref{sec:results_brownian}) all suggest that individual encounters leave a measurable imprint on the binary's orbital evolution. With only four realisations per scenario, we cannot unambiguously trace each eccentricity offset to a specific passage, nor pin down the precise value of the threshold across merger configurations; we assess the statistical significance and the sufficiency of the present suites in caveat~(i) below.

While our $N$-body simulations show that the $10^8 \msun$ perturbers are associated with measurable stochasticity in eccentricity, we note that they are modelled as softened extended particles ($\varepsilon\approx267\,\mathrm{pc}$ for the $10^{8}\msun$ case; Table~\ref{tab:particle_inventory_softening}) rather than point masses, with a compactness factor $\mathcal{C}=0.5$ that already renders them twice as concentrated as the standard mass-dependent prescription. For distant passages ($r\gg\varepsilon$) the perturber force is indistinguishable from that of a point mass, but the softening caps the peak impulse a perturber can deliver, $\Delta v_{\max}\sim G\massp/(\varepsilon V)$, and suppresses all closer encounters: at $r\sim\varepsilon$ a genuine point mass would exert roughly twice the force, while the hardest, smallest-impact-parameter passages ($b\lesssim\varepsilon$) that dominate single-encounter eccentricity kicks for a compact object are absent altogether. Because $\varepsilon\approx267\,\mathrm{pc}$ exceeds the binary scale at which the per-encounter kick maximises ($b\sim a$; Eq.~\eqref{eq:angmom_kick}), and is comparable to or larger than the physical sizes of real $10^{8}\msun$ compact systems (stripped nuclei and ultra-compact dwarfs, $r_{\rm eff}\sim$ tens of pc; IMBH-hosting nuclear star clusters, $r_{\rm eff}\sim$ few pc), this suppressed regime is physically realisable. The measured factor-of-$2.4$ enhancement in $\sigma_e$ is therefore a conservative lower bound on the physical near-impulsive effect for unresolved compact objects of the same mass, such as IMBHs in nuclear star clusters or stripped nuclei; we cannot quantify the magnitude of the missing contribution without simulations at finer force resolution.

Objects of such mass in the central regions of a merger remnant are a rare astrophysical occurrence. In massive elliptical galaxies, which are the primary hosts of PTA-relevant MBHBs, the population of perturbers is dominated by globular clusters with typical masses of $10^5\text{--}10^6 \msun$ \citep{Harris2015}, with only the most massive stripped galactic nuclei reaching $\sim 10^7\text{--}10^8 \msun$ \citep{Pfeffer2014, Antonini2015}. For the binary mass studied here these clusters correspond to $\mu_{\rm p}\sim 3\times10^{-5}\text{--}3\times10^{-4}$, i.e.\ single-encounter kicks $\Delta e_{\rm single}\sim[(1-e^2)/e]\,\mu_{\rm p}\sim10^{-5}\text{--}10^{-4}$ at the control median $e\approx0.64$, well below the noise floor; the same holds for IMBHs ($\massp\sim10^{5}\msun$, $\mu_{\rm p}\sim3\times10^{-5}$). Such objects therefore sit in the diffusive regime, and only the rare stripped-nucleus tail approaches the near-impulsive threshold $\mu_{\rm p}\gtrsim10^{-2}$ probed by our $10^{8}\msun$ runs. At fixed total perturber mass the diffusion coefficient scales as $D_e\propto\massp$ (Section~\ref{sec:theory}), so redistributing the same mass budget into a larger number of lower-mass objects reduces the diffusive contribution rather than enhancing it; unresolved low-mass populations are therefore less effective at randomising the eccentricity, not more, unless their mass function is far steeper or more centrally concentrated than observed. Even at the upper end of this range, for such an object to deliver a dynamically significant kick it must pass within a few binary separations (the threshold scales as $[(1-e^2)/e]^{-1}$; Eq.~\eqref{eq:angmom_kick}) during the short interval over which the formation eccentricity is set, a condition unlikely to be met given the rarity of such objects and the short duration of the formation window. Our results therefore suggest that, in realistic environments, the effective perturber mass scale lies below the near-impulsive regime, and the associated eccentricity stochasticity remains below the numerical noise floor of our current $N$-body simulations. Impulsive perturber--BH encounters are thus unlikely to be a dynamically relevant pathway for randomising eccentricity at binary formation for the majority of MBHB mergers of interest to PTA detection.

Although our simulations adopt a single binary mass ($\massb \approx 3.1\times10^{9}\,\msun$), the relevant control parameter is the dimensionless ratio $\mu_{\rm p} = \massp/\massb$ rather than $\massp$ and $\massb$ separately: the single-encounter kick $\Delta e_{\rm single} \sim [(1-e^2)/e]\,\mu_{\rm p}$ scales directly with the individual perturber--binary mass ratio at fixed orbital geometry, while the diffusion coefficient $D_e \propto \np\massp^{2}$ depends in addition on the perturber number density and spatial distribution. The diffusive--near-impulsive transition we identify therefore maps across the PTA-relevant mass range ($\massb \sim 10^{8}$--$10^{10}\,\msun$; \citealp{sesana_2013, agazie_2023}) under a rescaling of $\mu_{\rm p}$. A fixed perturber mass acts more impulsively on a lighter binary: a $10^{7}\,\msun$ perturber and a $10^{8}\,\msun$ binary give $\mu_{\rm p}=0.1$, larger than the near-impulsive ratio of our $10^{8}\,\msun$ suite, so such a configuration would be expected to randomise the formation eccentricity even more strongly. This is consistent with the broad spread we already measure at $\mu_{\rm p}\approx0.03$ ($e_{\rm b}\approx0.30$--$0.99$), and with the still-wider spread anticipated at $\mu_{\rm p}\sim0.1$. Whether the conclusion drawn at $\massb\approx10^{9}\,\msun$ generalises across the whole PTA band depends on how the perturber population scales with host mass. If the dominant perturbers, globular clusters and stripped nuclei, track the host stellar mass through an approximately fixed mass fraction, as in the bulge-fraction scaling adopted here, their characteristic $\mu_{\rm p}$ remains roughly constant with $\massb$, and the dominant population stays on the diffusive side of the transition across the entire range. The exception is the rare massive tail at the low-mass end of the band: for $\massb\sim10^{8}\,\msun$ a single stripped nucleus of $\sim10^{7}\,\msun$ already reaches $\mu_{\rm p}\sim0.1$ and can act impulsively, so perturber-driven randomisation may be comparatively more relevant for the lowest-mass PTA binaries than for the most massive ones.

The $10^{7}\msun$ runs show a median formation eccentricity $\langle e_{\rm b}\rangle\approx 0.41$, lower than the control ($\approx 0.64$). The scatter-based diagnostics on which our conclusions rest are unaffected by this offset: the paired shifts (Fig.~\ref{fig:formation_summary}d) are modest, $\sigma_e$ is indistinguishable from the control, and no accompanying dynamical signatures appear in Section~\ref{sec:results_dynamics}. We therefore do not draw a physical conclusion from this median offset given the four-realisation sample. Additional seeds would primarily sharpen this secondary median-shift question, since the scatter, paired shifts, and dynamical diagnostics all place the $10^{7}\msun$ suite at the no-perturber noise floor.

We note several caveats. (i) \textit{Statistical power}: with four realisations per scenario we do not aim to measure a precise population-level variance, nor do we base our conclusion on $\sigma_e$ alone. Treating the estimated noise floor as exact would overstate the significance, since $\sigma_e\approx0.11$ is itself measured from only four control realisations; the appropriate two-sample $F$-test of the variances gives $F=(0.26/0.11)^2\approx5.6$ with $(3,3)$ degrees of freedom, a one-sided $p\approx0.1$ (two-sided $p\approx0.2$); we quote $\sigma_e$ as the population standard deviation, and because the two suites contain equal numbers of realisations the variance ratio is unchanged under the Bessel-corrected (sample-variance) convention. Our conclusion instead rests on the variance enhancement being accompanied by distinct dynamical diagnostics (event-aligned residuals in orbital energy and angular momentum, torque spikes, and the enhanced centre-of-mass wander), all weaker or absent in the control and $10^{7}\msun$ runs, which a pure $\sigma_e$ fluctuation would not reproduce. This combination is sufficient to motivate the qualitative diffusive-to-near-impulsive distinction; a larger ensemble would refine the precise threshold in $\mu_{\rm p}$ rather than alter it. (ii) \textit{Single merger configuration}: we explored only one initial orbital configuration ($e_{\rm orb}=0.987$, near-radial). Different mass ratios,
inclinations, or orbital eccentricities may yield a different threshold
$\mu_{\rm p}$, especially through the eccentricity-dependent prefactor
$(1-e^2)/e$. (iii) \textit{Single-mass perturber populations}: we adopted equal-mass perturbers at only two mass scales ($10^{7}\msun$, $10^{8}\msun$), chosen to probe
either side of the diffusive--impulsive transition rather than to reproduce
a realistic mass distribution. More $N$-body runs at intermediate masses would help locate the precise threshold in $\mu_{\rm p}$, but would not affect the main conclusions, since the realistic perturber population sits below the transition. (iv) \textit{Idealised initial conditions}: although the model parameters are drawn from a TNG100-1 major merger, the galaxies are realised as smooth, spherical Hernquist bulge and halo components rather than the full cosmological progenitor. The absence of triaxiality, streams, and other cosmological substructure is likely to make our setup conservative: a lumpier or non-spherical potential would introduce additional stochasticity, so the measured $\sigma_e$ values are plausibly lower bounds on the scatter expected in a fully cosmological environment. We caution, however, that a realistic potential need not act purely as added noise: coherent triaxial torques could also alter the inspiral and formation time, so this is the likely rather than guaranteed direction. (v) \textit{Neglected perturber segregation}: both perturber suites are initialised from the bulge Hernquist profile, so we do not impose the greater central concentration that heavier perturbers would inherit from prior dynamical friction. As argued in Section~\ref{sec:methodology}, this is also conservative for the $10^{8}\msun$ case, since a self-consistent treatment would place more massive perturbers closer to the binary at $t_{\rm b}$. (vi) \textit{Neglected gas}: the progenitors are modelled as collisionless, yet the primary retains a star-forming gas reservoir ($f_{\rm gas}\approx0.19$ within its stellar half-mass radius; Section~\ref{sec:methodology}), and we do not treat its omission as conservative, since gas acts in two competing directions. A gas-rich nucleus fragments into giant molecular clouds and, at high redshift, clumps of up to $10^{8}$--$10^{9}\msun$ \citep{Dekel2009, Genzel2011}; by the $\mu_{\rm p}$ criterion of Section~\ref{sec:theory} these add a perturber population of comparable total mass that could act near-impulsively and raise $\sigma_e$. Dissipation acts oppositely: drag on the inspiralling MBHs \citep{Mayer2007} and, once a circumbinary disc forms, evolution towards an equilibrium eccentricity \citep{Roedig2011, DOrazioDuffell2021} regularise the orbit and lower $\sigma_e$. Which dominates cannot be settled without hydrodynamical re-simulations, the main extension needed to generalise to gas-rich mergers.

The numerical noise floor is set by the mass resolution of the galactic nucleus: \citet{gualandris2026converging} find $\sigma_e \propto N^{-1/2}$, so it falls only as the square root of the computational cost, and at our resolution ($N_{\rm bulge}\approx 2\times10^6$) it reproduces the measured $\sigma_e\approx 0.11$. Resolving $e_{\rm b}$ to few-per-cent precision ($\sigma_e\sim 0.03$) would require $\sim(0.11/0.03)^2\approx 13$ times more bulge particles, or $\approx 2.6\times10^7$. Even this leaves the merger time uncertain at the tens-of-per-cent level: the gravitational-wave inspiral time is a steep function of eccentricity, $t_{\rm GW}\propto(1-e^2)^{7/2}$ \citep{Peters1964}, so $|\mathrm{d}\ln t_{\rm GW}/\mathrm{d}e| = 7e/(1-e^2)\approx 7.6$ at the control median $e\approx0.64$, and a residual $\sigma_e\sim0.03$ still propagates to $\delta\ln t_{\rm GW}\approx 7.6\times0.03\approx 0.2$. A single such run is tractable with \texttt{Griffin}'s approximately $\mathcal{O}(N)$ FMM scheme, but the matched suite needed to measure $\sigma_e$ multiplies the cost in proportion, placing few-per-cent convergence beyond the present study; adding seeds at fixed resolution reduces the sampling error but not the floor itself. This estimate is approximate, as \citet{gualandris2026converging} adopt single-component models whereas ours are multi-component.

\section{Conclusions}\label{sec:conclusions}

We have presented high-resolution $N$-body re-simulations of a major merger from IllustrisTNG100-1 to test whether massive perturbers can induce physical stochasticity in the formation eccentricity of MBHBs. Using the \texttt{Griffin} code, we evolved a no-perturber control and two matched perturber suites with $10^7\msun$ and $10^8\msun$ perturbers, each with four independent realisations. Both perturber suites redistribute a generous fraction $f_{\rm target}=0.1$ of the primary bulge mass into perturbers ($M_{\rm total}\approx3.16\times10^{10}\msun$); relative to the median smooth-bulge particle mass, these perturbers are approximately $47$ and $465$ times more massive than the background stellar particles, respectively (Table~\ref{tab:particle_inventory_softening}).

The no-perturber control yields $\sigma_e \approx 0.11$ at a median formation eccentricity $\langle e_{\rm b}\rangle \approx 0.64$, consistent in normalisation with the empirical Poisson noise floor expected at the adopted resolution. The $10^7\msun$ suite ($\mu_{\rm p}\approx3.2\times10^{-3}$) gives $\sigma_e\approx0.115$, indistinguishable from the control. This agrees with the binary--single scattering estimate $\Delta e_{\rm single}\sim[(1-e^2)/e]\,\mu_{\rm p}\sim0.003$, which places individual $10^7\msun$ encounters below the numerical noise floor and in the diffusive regime.

The $10^8\msun$ suite ($\mu_{\rm p}\approx3.2\times10^{-2}$) instead yields $\sigma_e\approx0.26$, a factor of $\approx2.4$ above the control, with formation eccentricities spanning $e_{\rm b}\approx0.30$--$0.99$. The same runs show larger event-aligned residuals in orbital energy and angular momentum, together with stronger torque spikes, supporting the interpretation that rare near-impulsive perturber--MBHB encounters, occurring as the binary forms and its angular momentum is set, can leave a measurable imprint on the formation eccentricity rather than on its subsequent hardening. Event-by-event confirmation would require either larger ensembles or targeted encounter-isolation experiments, but the present matched suites probe a transition consistent with the expected diffusive-to-near-impulsive scaling, with the enhanced scatter accompanied by corroborating, physically related energy, angular-momentum, torque, and COM-wander signatures.

The astrophysically expected perturber population in massive elliptical galaxies, dominated by globular clusters and stripped nuclei with characteristic masses $\lesssim10^7\msun$, lies on the diffusive side of this transition. We therefore conclude that perturber-driven eccentricity randomisation is unlikely to be a significant pathway for GWB-relevant MBHB mergers in massive ellipticals. This conclusion rests on the mass-ratio margin rather than on the statistically marginal $10^8\msun$ enhancement: the dominant realistic perturber population falls one to two orders of magnitude below the transition in $\mu_{\rm p}$ (with only the rare stripped-nucleus tail approaching it), so the conclusion holds whether or not the measured factor-of-$2.4$ excess is significant. Larger simulation suites spanning merger geometry and a continuous perturber mass spectrum would refine the threshold in $\mu_{\rm p}$, rather than alter the main astrophysical conclusion that the dominant perturber population expected in massive ellipticals lies below the near-impulsive regime.

\section*{Acknowledgements}
AG acknowledges support from grant ST/Y002385/1.

%%%%%%%%%%%%%%%%%%%%%%%%%%%%%%%%%%%%%%%%%%%%%%%%%%
\section*{Data Availability}
The data underlying this article will be shared on reasonable request to the corresponding author. The analysis scripts used to process the simulation outputs and generate the figures will also be shared on reasonable request. This work made use of the \texttt{Griffin} $N$-body code and standard Python scientific libraries, including \texttt{NumPy}, \texttt{SciPy}, \texttt{pandas}, \texttt{h5py}, \texttt{Matplotlib}, and \texttt{SciencePlots}.


\bibliographystyle{mnras}
\bibliography{references} % if your bibtex file is called example.bib

%\begin{thebibliography}{99}
%\end{thebibliography}

%%%%%%%%%%%%%%%%%%%%%%%%%%%%%%%%%%%%%%%%%%%%%%%%%%

% Don't change these lines
\bsp	% typesetting comment
\label{lastpage}
\end{document}